\documentclass[aps,prl,twocolumn,tightenlines,superscriptaddress,showpacs,byrevtex]{revtex4}

\usepackage{graphicx}
\usepackage{dcolumn}
\usepackage{bm}

\graphicspath{{./}}

\def\DE {\Delta E}
\def\Mbc {M_{\rm bc}}

\begin{document}
\title{
 Search for $B^{+} \to J/\psi \eta' K^{+}$ and
 $B^{0} \to J/\psi \eta' K_S^0$ Decays }

\affiliation{Budker Institute of Nuclear Physics, Novosibirsk}
\affiliation{Chiba University, Chiba}
\affiliation{Chonnam National
University, Kwangju}
\affiliation{University of Cincinnati,
Cincinnati, Ohio 45221}
\affiliation{Department of Physics, Fu Jen
Catholic University, Taipei}
\affiliation{The Graduate University for Advanced Studies, Hayama, Japan} 
\affiliation{University of Hawaii, Honolulu, Hawaii 96822}
\affiliation{High Energy Accelerator Research Organization (KEK),
Tsukuba} \affiliation{Hiroshima Institute of Technology, Hiroshima}
\affiliation{Institute of High Energy Physics, Chinese Academy of
Sciences, Beijing} \affiliation{Institute of High Energy Physics,
Vienna} \affiliation{Institute of High Energy Physics, Protvino}
\affiliation{Institute for Theoretical and Experimental Physics,
Moscow} \affiliation{J. Stefan Institute, Ljubljana}
\affiliation{Kanagawa University, Yokohama} \affiliation{Korea
University, Seoul} \affiliation{Kyungpook National University,
Taegu} \affiliation{Swiss Federal Institute of Technology of
Lausanne, EPFL, Lausanne} \affiliation{University of Ljubljana,
Ljubljana} \affiliation{University of Maribor, Maribor}
\affiliation{University of Melbourne, Victoria} \affiliation{Nagoya
University, Nagoya} \affiliation{Nara Women's University, Nara}
\affiliation{National Central University, Chung-li}
\affiliation{National United University, Miao Li}
\affiliation{Department of Physics, National Taiwan University,
Taipei} \affiliation{H. Niewodniczanski Institute of Nuclear
Physics, Krakow} \affiliation{Nippon Dental University, Niigata}
\affiliation{Niigata University, Niigata} \affiliation{University of
Nova Gorica, Nova Gorica} \affiliation{Osaka City University, Osaka}
\affiliation{Osaka University, Osaka} \affiliation{Panjab
University, Chandigarh} \affiliation{Peking University, Beijing}
\affiliation{University of Pittsburgh, Pittsburgh, Pennsylvania
15260} \affiliation{RIKEN BNL Research Center, Upton, New York
11973} \affiliation{University of Science and Technology of China,
Hefei} \affiliation{Seoul National University, Seoul}
\affiliation{Shinshu University, Nagano} \affiliation{Sungkyunkwan
University, Suwon} \affiliation{University of Sydney, Sydney NSW}
\affiliation{Tata Institute of Fundamental Research, Bombay}
\affiliation{Toho University, Funabashi} \affiliation{Tohoku Gakuin
University, Tagajo} \affiliation{Tohoku University, Sendai}
\affiliation{Department of Physics, University of Tokyo, Tokyo}
\affiliation{Tokyo Institute of Technology, Tokyo}
\affiliation{Tokyo Metropolitan University, Tokyo}
\affiliation{Tokyo University of Agriculture and Technology, Tokyo}
\affiliation{Virginia Polytechnic Institute and State University,
Blacksburg, Virginia 24061} \affiliation{Yonsei University, Seoul}
  \author{Q.~L.~Xie}\affiliation{Institute of High Energy Physics, Chinese Academy of Sciences, Beijing} 
  \author{K.~Abe}\affiliation{High Energy Accelerator Research Organization (KEK), Tsukuba} 
  \author{I.~Adachi}\affiliation{High Energy Accelerator Research Organization (KEK), Tsukuba} 
  \author{H.~Aihara}\affiliation{Department of Physics, University of Tokyo, Tokyo} 
  \author{D.~Anipko}\affiliation{Budker Institute of Nuclear Physics, Novosibirsk} 
  \author{K.~Arinstein}\affiliation{Budker Institute of Nuclear Physics, Novosibirsk} 
  \author{V.~Aulchenko}\affiliation{Budker Institute of Nuclear Physics, Novosibirsk} 
  \author{T.~Aushev}\affiliation{Swiss Federal Institute of Technology of Lausanne, EPFL, Lausanne}\affiliation{Institute for Theoretical and Experimental Physics, Moscow} 
  \author{T.~Aziz}\affiliation{Tata Institute of Fundamental Research, Bombay} 
  \author{S.~Bahinipati}\affiliation{University of Cincinnati, Cincinnati, Ohio 45221} 
% ...
\collaboration{The Belle Collaboration}

\date{\textbf{\today}}

\begin{abstract}
We report the results of searches for $B^{+} \to J/\psi \eta' K^+$
and $B^{0} \to J/\psi \eta' K_S^0$ decays, using a sample of 388
$\times 10^6$ $B\bar{B}$ pairs collected with the Belle detector at
the $\Upsilon(4S)$ resonance. No statistically significant signal is
found for either of the two decay modes and upper limits for the
branching fractions are determined to be $\mathcal{B}(B^{+} \to
J/\psi \eta' K^+) < 8.8 \times 10^{-5}$ and $\mathcal{B}(B^{0} \to
J/\psi \eta' K_S^0) < 2.5 \times 10^{-5}$ at 90\% confidence level.
\end{abstract}

 \maketitle

Studies of exclusive $B$ meson decays to charmonium play an
important role in exploring $CP$ violation \cite{charmonium_cpv} and
in observations of new resonant states that include a $(c\bar c)$
pair \cite{new_state} \cite{new_state2}. The decay $B \to J/\psi
\eta' K$ proceeds via a Cabibbo-allowed and color-suppressed
transition ($b \to c\bar cs$) with $(s\bar s)$ quark popping. A
branching fraction comparable to that of the decay $B \to J/\psi
\phi K$ \cite{jpsiphi} is therefore expected.

The $B \to J/\psi \eta' K$ decay modes are of particular interest in
the search for hybrid charmonium states, $\psi_g$, which are excited
gluonic $(c\bar c g)$ states, which may decay to $J/\psi
+(\eta^{(\prime)},\phi)$ \cite{hybrid}. These hybrid states are
expected to be produced in $B$ decays such as $B \to \psi_g K$. A
near-threshold $\omega$ $J/\psi$ enhancement in $B \to K \omega
J/\psi$ decays observed by Belle \cite{new_state2} has some
properties similar to those predicted for $c\bar c$-gluon hybrid
states \cite{hybrid}. A state, $Y(4260)$, recently observed by BaBar
 in $e^+e^- \to Y(4260) \gamma_{\rm (ISR)}$ transitions
\cite{y4260}, is also a candidate for such a $\psi_g$ state
\cite{close06}, where $\gamma_{\rm (ISR)}$ denotes an initial state
radiation photon. An enhanced decay rate to $J/\psi \eta^{(\prime)}$
modes would be a supporting evidence for this assignment.

In this paper, we report results on searches for the decay modes
$B^{+} \to J/\psi \eta' K^+$ and $B^{0} \to J/\psi \eta' K_S^0$
\cite{CC} in a sample of 357 $\rm fb^{-1}$ containing 388 $\times
10^6$ $B\bar{B}$ pairs accumulated at the $\Upsilon$(4$S$) resonance
with the Belle detector \cite{belle} at the KEKB energy asymmetric
$e^+e^-$ collider \cite{kekb}.  These are the first results ever
presented for these decay modes.

The Belle detector is a large-solid-angle magnetic spectrometer that
consists of a silicon vertex detector, a 50-layer central drift
chamber (CDC), an array of aerogel threshold \v{C}erenkov counters
(ACC), a barrel-like arrangement of time-of-flight scintillation
counters (TOF), and an electromagnetic calorimeter comprised of
CsI(Tl) crystals (ECL). These detectors are located inside a
superconducting solenoid coil that provides a 1.5 T magnetic field.
An iron flux-return located outside of the coil is instrumented to
detect $K_L^0$ mesons and to identify muons.

Candidates for these two decay modes are reconstructed with the
decay chains $J/\psi \to l^+ l^-$ ($l = e, \mu$), $\eta ' \to \eta
\pi^+ \pi^-$, $\eta \to \gamma \gamma$ and $K_S^0 \to \pi^+ \pi^-$.

Events are required to pass a basic hadronic event selection
\cite{PRD03}. To suppress continuum background ($e^+e^- \to q \bar
q$, where $q = u, d, s, c$), we require the ratio of the second to
zeroth Fox-Wolfram moments \cite{FW} to be less than 0.5 and the
absolute value of the cosine of the angle between the $B$ meson and
beam direction in the center-of-mass system to be less than 0.85.

The selection criteria for the $J/\psi$ decaying to $l^+l^-$ are
identical to those used in our previous papers \cite{PRD03}.
 $J/\psi$ candidates are pairs of oppositely
charged tracks that originate from a region within 5~cm of the
nominal interaction point (IP) along the beam direction and are
positively identified as leptons. In order to reduce the effect of
bremsstrahlung or final state radiation, photons detected in the ECL
within $0.05$ radians of the original $e^-$ or $e^+$ direction are
included in the calculation of the $e^+e^-(\gamma)$ invariant mass.
Because of the radiative low-mass tail, the $J/\psi$ candidates are
required to be within an asymmetric invariant mass window:
$-150(-60)$ $\rm{MeV}$/$c^2$
$<M_{e^+e^-(\gamma)}(M_{\mu^+\mu^-})-m_{J/\psi} <$+36(+36)
$\rm{MeV}$/$c^2$, where $m_{J/\psi}$ is the nominal $J/\psi$ mass
\cite{pdg}. In order to improve the momentum resolution of the
$J/\psi$ signal, a vertex and mass constrained fit to the
reconstructed $J/\psi$ candidates is then performed and a loose cut
on the vertex fit quality is applied.

In order to identify hadrons, a likelihood $L_i$ for each  hadron
type $i$ ($i = \pi, K$ and $p$) is formed using information from the
ACC, TOF and CDC ($dE/dx$). Charged tracks that were previously
identified as electrons or muons are rejected in the hadron
identification procedure. The kaon from the $B$ meson and the pions
from $\eta '$ are selected with the requirements of
$L_{K}/(L_{K}+L_{\pi})> 0.4$ and $L_{\pi}/(L_{\pi}+L_{K})> 0.1$,
which have efficiencies of 82.5\% and 89.4\%, respectively.

The $\eta$ from $\eta '$ decay is reconstructed by combining two
photons that do not match to any charged track. The calorimeter
cluster energy, $E_{\gamma}$, is required to exceed $50$ $\rm{MeV}$
for both photons.  To veto photons from $\pi^{0}$, we combine each
photon candidate with another photon (with $E_{\gamma}>50$
$\rm{MeV}$) and reject it if the invariant mass is consistent with a
$\pi^{0}$: $|M_{\gamma\gamma} - m_{\pi^{0}}| < 18$ $\rm{MeV}/c^{2}$.
For $\eta$ candidates, $\cos\theta^{\eta}_{\rm hel}$ is used to
suppress background, where $\theta^{\eta}_{\rm hel}$ is defined as
the angle between the photon momentum in the $\eta$ rest frame and
the boost direction of $\eta$ in the laboratory frame. Signal events
have a uniform $\cos\theta^{\eta}_{\rm hel}$ distribution, whereas
random photon pairs have a distribution that peaks near $\pm 1$. We
require $|\cos\theta^{\eta}_{\rm hel}|<0.94$ (0.97) for charged
(neutral) $B$ decay. The invariant mass of the $\eta$ candidate is
required to be within $0.496 ~{\rm MeV}/c^{2} < M_{\gamma\gamma} <
0.582 ~{\rm MeV}/c^{2}$ ($3\sigma$ of the nominal $\eta$ mass). To
improve the momentum resolution of the $\eta$, we apply a mass
constrained fit to the reconstructed $\eta$.

We reconstruct $\eta'$ candidates by combining the $\eta$ and
selected $\pi^+$¡¢$\pi^-$ candidates. We require the invariant mass
to be $0.940 ~{\rm MeV}/c^{2} < M_{\eta \pi^+ \pi^-} < 0.975 ~{\rm
MeV}/c^ {2}$. To improve the momentum resolution, we apply a vertex
and mass constrained fit on the $\eta'$ \cite{plb01}.

For $K_S^0$ candidates, we impose momentum-dependent requirements on
the impact parameters from the IP for both $K_S^0$ daughter tracks, the distance
between the daughter tracks along the beam axis at the $K_S^0$
vertex, the difference of azimuthal angles between the $K_S^0$
momentum and the direction of the $K_S^0$ vertex from the IP, and
the flight length of the $K_S^0$. The invariant mass of the $K_S^0$
candidate is required to be within 16~$\rm{MeV}/c^2$ ($3\sigma$) of
the $K_S^0$ mass. We apply vertex and mass constrained fits for the
$K_S^0$ candidates to improve the momentum resolution.

These criteria maximize $N_{\rm sig}/\sqrt{N_{\rm sig}+N_{\rm
bkg}}$, where $N_{\rm sig}$ is the number of expected signal events
from signal Monte Carlo (MC) samples with assumed branching
fractions of $1.0 \times 10^{-5}$ for both $B^{+} \to J/\psi \eta '
K^+$ and $B^{0} \to J/\psi \eta ' K_S^0$ modes, and $N_{\rm bkg}$ is
the number of expected background events estimated from a MC sample
of $B\bar B$-pairs with a $J/\psi$ in the final state.

$B$ mesons are reconstructed by combining a $J/\psi$, an $\eta '$
and a $K$ candidate. We identify $B$ candidates using two widely
used kinematic variables calculated in the center-of-mass system:
the beam-energy constrained mass ($\Mbc \equiv \sqrt{E_{\rm
beam}^{2}-P_{B}^{2}}$) and the energy difference ($\DE \equiv E_{B}-
E_{\rm beam}$), where $E_{\rm{beam}}$ is the beam energy, $P_{B}$
and $E_{B}$ are the reconstructed momentum and energy of the $B$
candidate. We select both $B^+$ and $B^0$ candidates within the
range $|\DE|<0.20$ $\rm{GeV}$ and $\Mbc >5.20$ $\rm{GeV}/c^{2}$ for
the final analysis. The signal regions are defined to be $5.27$
GeV/$c^2$ $< \Mbc < 5.29$ GeV/$c^2$ and $|\DE| <$ 0.03 ${\rm GeV}$
for both $B^{+} \to J/\psi \eta ' K^{+}$ and $B^{0} \to J/\psi \eta
' K_S^0$ modes, which corresponds to three standard deviations based
on the MC simulation.

After all selection requirements, about 26.5\% of the $B^{+} \to
J/\psi \eta ' K^{+}$ candidates have more than one entry per event;
this occurs for 30.0\% of the $B^{0} \to J/\psi \eta ' K_S^0$
candidates. For these events, the $B$ candidate with the best vertex
fit quality is used. From a study of MC simulated events, we
conclude that around 81.5\% of multiple $B^{+} \to J/\psi \eta '
K^{+}$ candidates and 70.0\% of multiple $B^{0} \to J/\psi \eta '
K_S^0$ candidates are selected correctly by this procedure.

The signal yields are extracted by maximizing the two dimensional
extended likelihood function,
\[
{\cal L} = \frac{e^{-\sum
\limits_{k}N_{k}}}{N!}\prod^N_{i=1}\left[\sum_{k}N_{ k}\times
P_k(M_{{\rm bc}}^{i},\Delta E^{i})\right],
\]
where $N$ is the total number of candidate events, $i$ is the
identifier of the $i$-th event, $N_{k}$ and $P_{k}$ are the yield
and probability density function (PDF) of the component $k$, which
corresponds to the signal and background.

The signal PDFs for the two decay modes are modeled using a sum of
two Gaussians for $\Mbc$ and a sum of two Gaussians plus a
bifurcated Gaussian that describes the tail of the distribution in
$\DE$. The PDF parameters are initially determined using signal MC
and subsequently the primary Gaussian parameters are corrected using
a control data sample of $B^+ \to J/\psi K^{*+}$ decays with $K^{*+}
\to \pi^0 K^+$. The signal shape parameters are kept fixed in the
fits to the data.

The dominant background comes from random combination of $J/\psi$,
$\eta'$, and $K$ candidates in $B\bar B$ events. The background from
continuum is found to be small (a few \%). A threshold function
\cite{ARGUS} is used for the $\Mbc$ PDF. For $\DE$, we use a
first-order polynomial function. The MC study shows that the
background from the decays with similar topology that would make a
peak in $\Mbc$ or $\DE$ distributions is negligible.

\begin{figure*}[hbtp]
\includegraphics[width=0.3\textwidth]{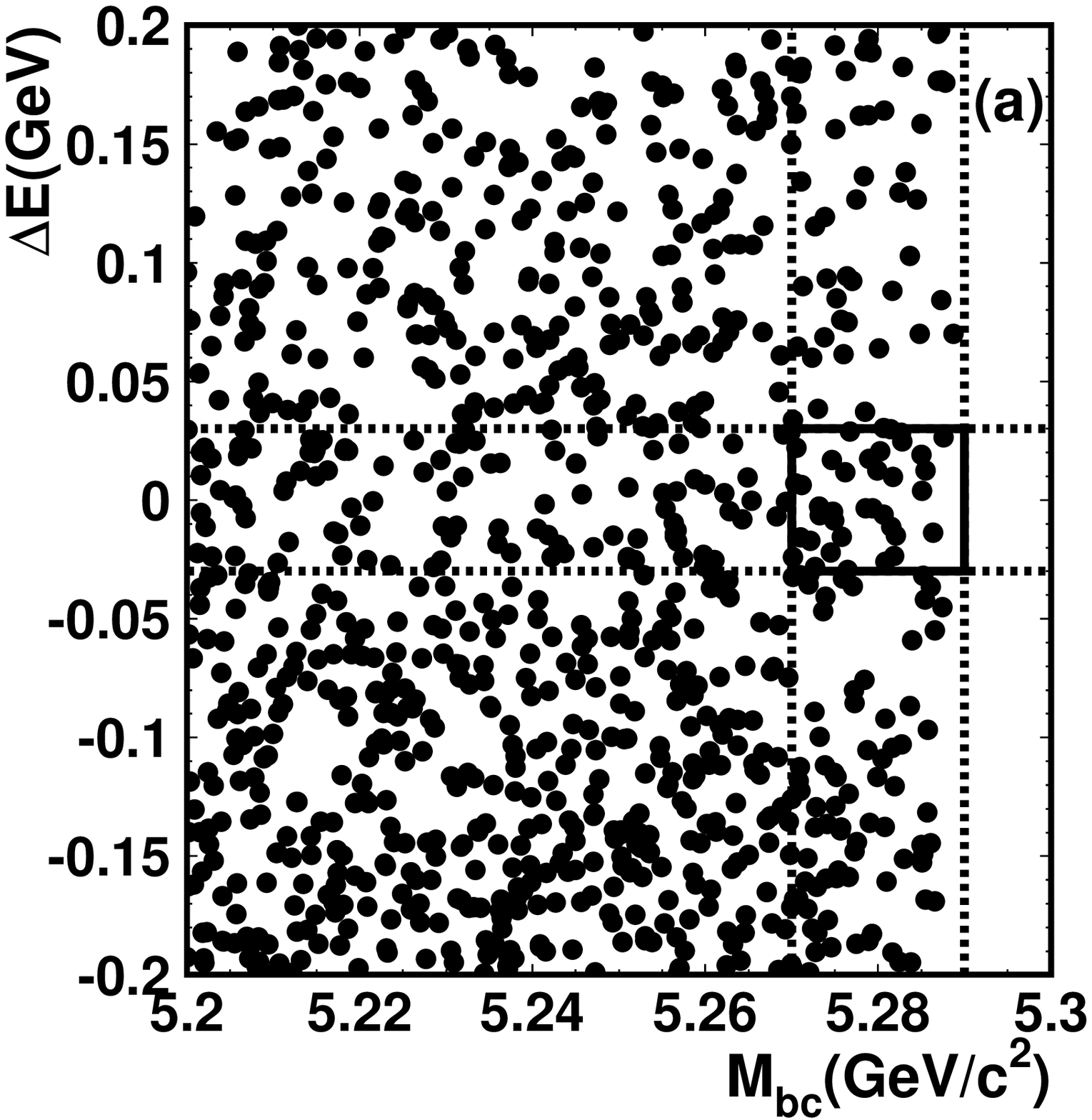}
\includegraphics[width=0.3\textwidth]{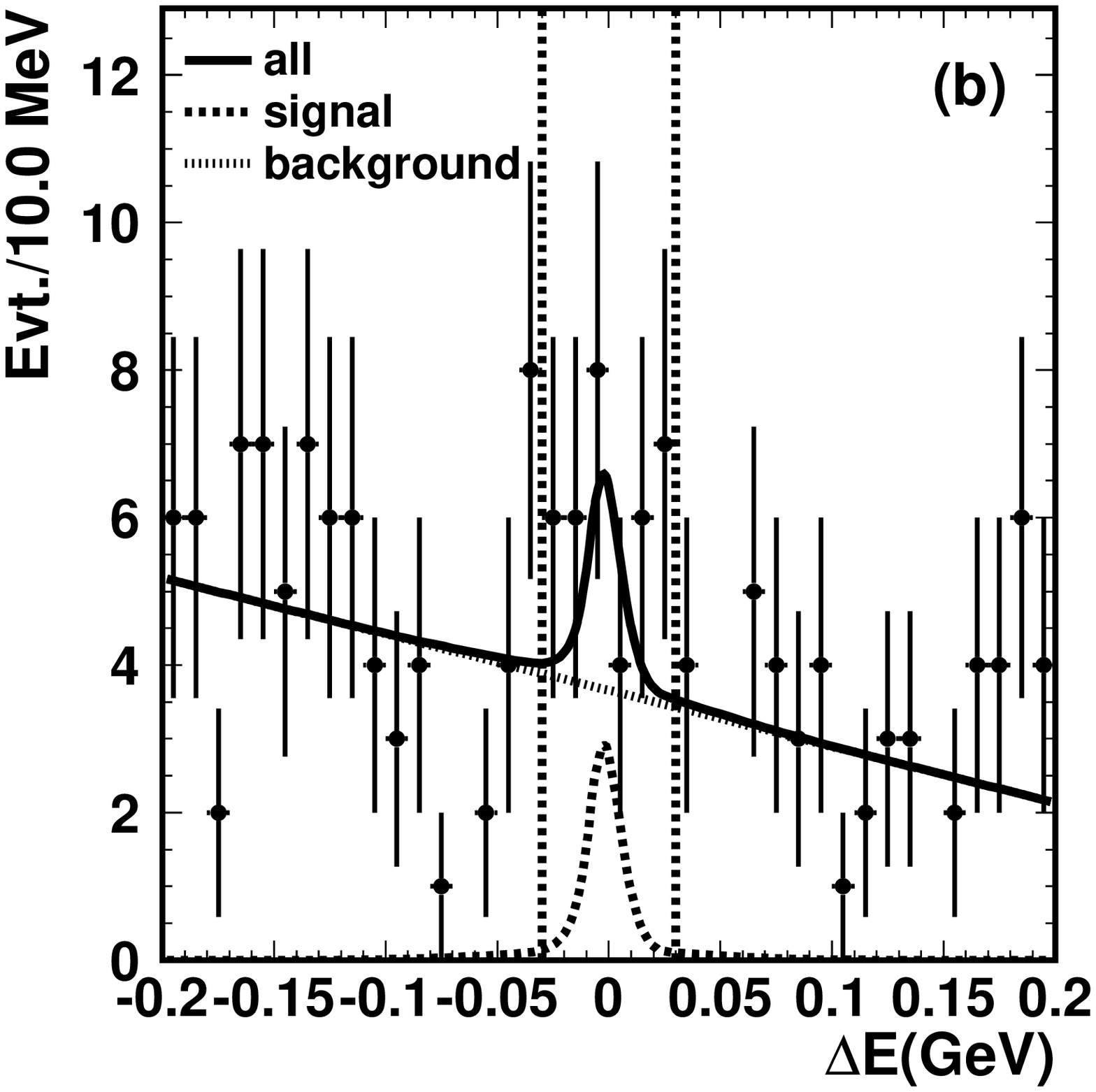}
\includegraphics[width=0.3\textwidth]{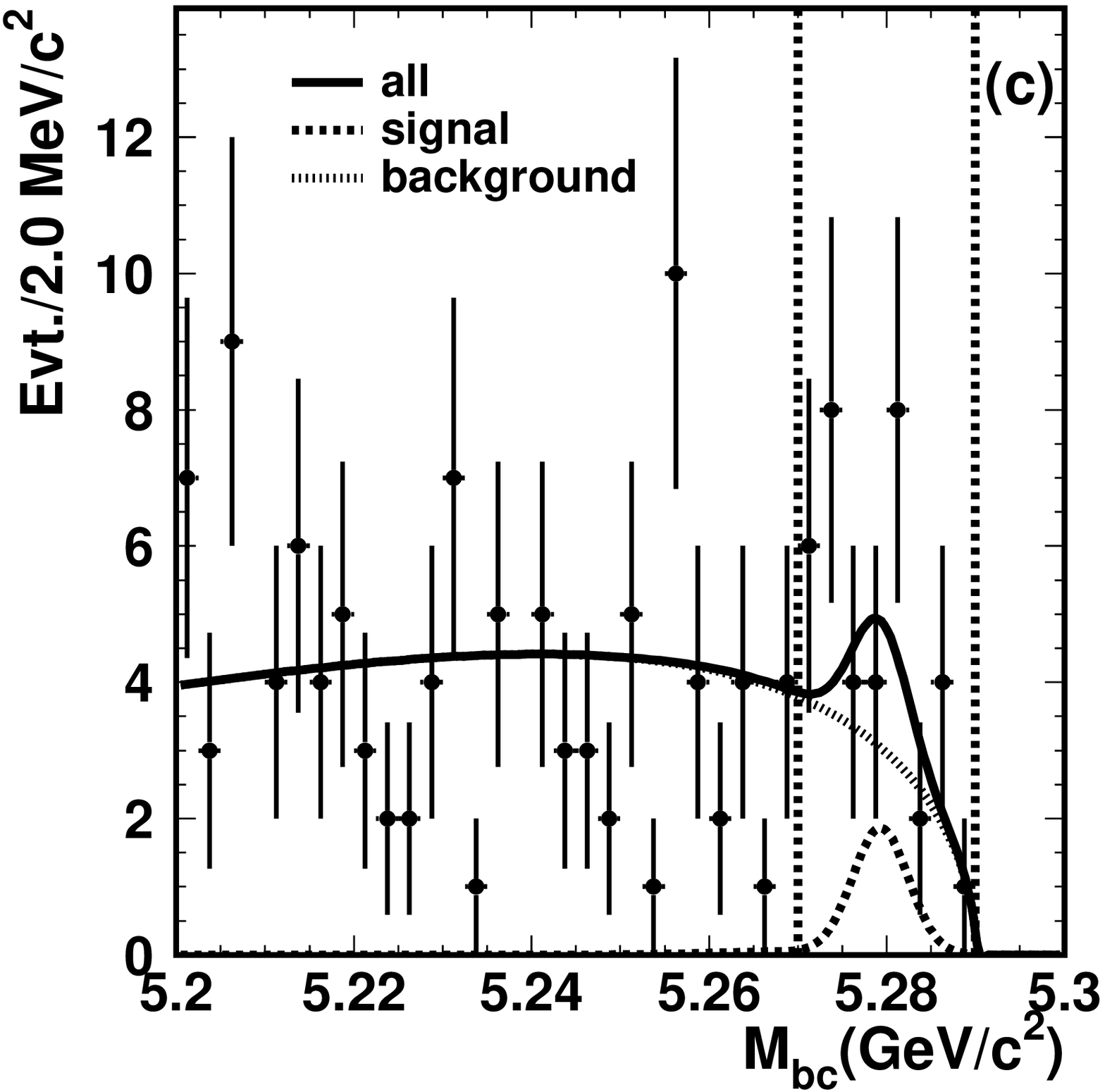}
\caption{ (a) The ($\Mbc$, $\DE$) scatterplot of $B^{+} \to J/\psi
\eta' K^{+}$ candidates and its projections onto (b) $\DE$ with 5.27
GeV/$c^2$ $< \Mbc <$ 5.29 GeV/$c^2$ and (c) $\Mbc$ with $ |\DE| <
0.03$ GeV. The dashed lines and solid boxes indicate the signal
regions. The curves are the result of the fit as described in the
text. } \label{sbox}
\end{figure*}

\begin{figure*}[htp]
\includegraphics[width=0.3\textwidth]{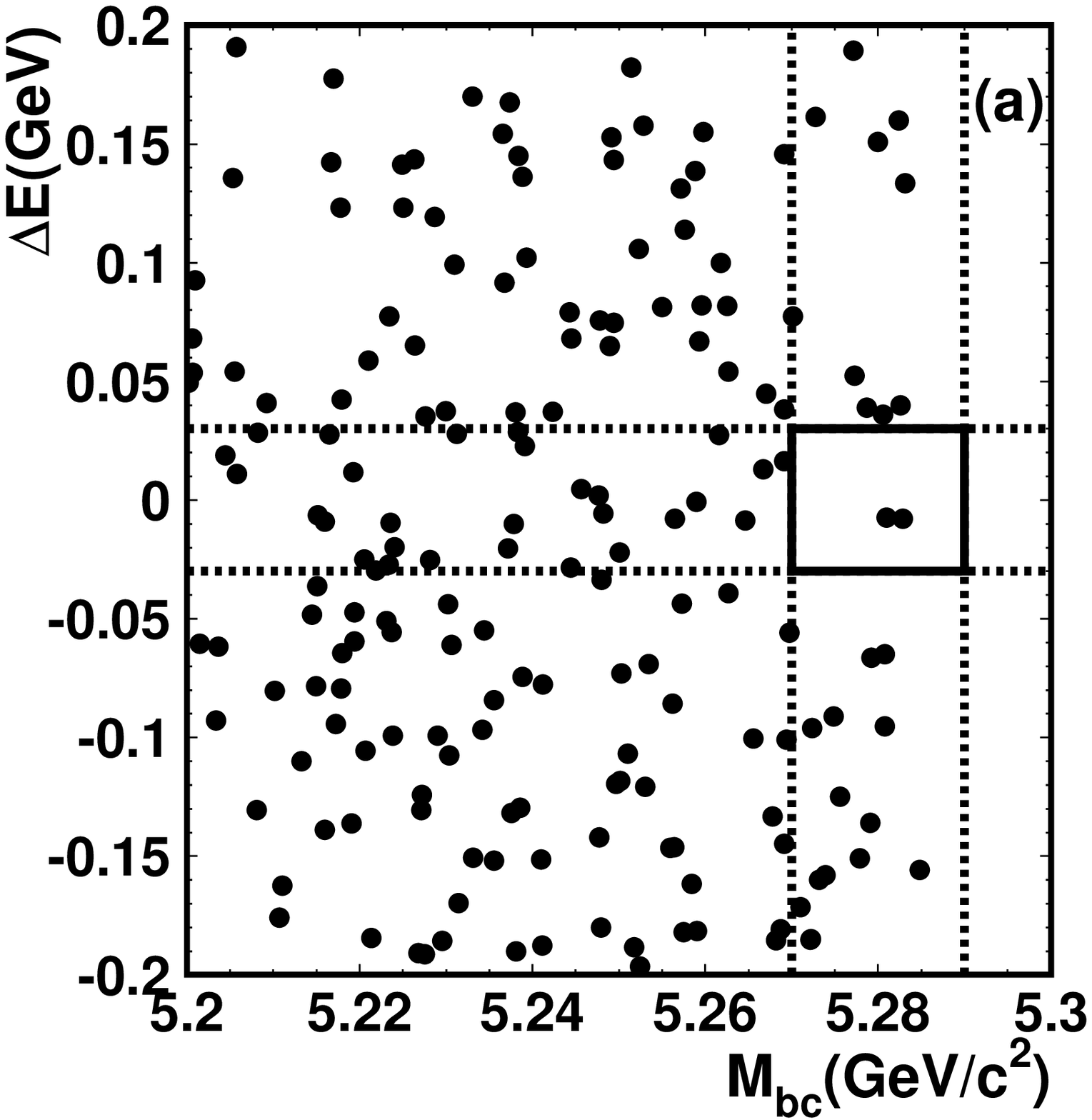}
\includegraphics[width=0.3\textwidth]{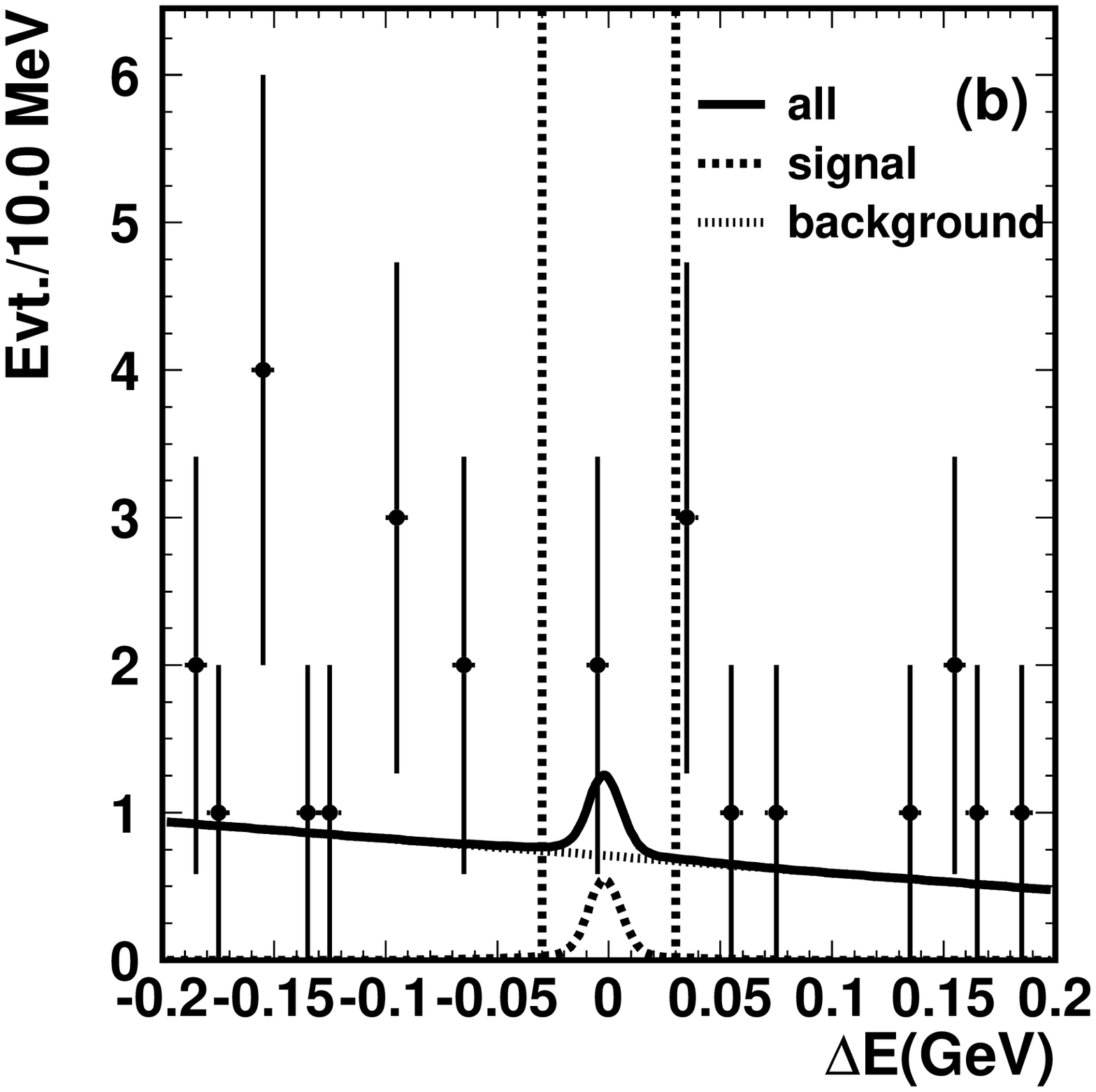}
\includegraphics[width=0.3\textwidth]{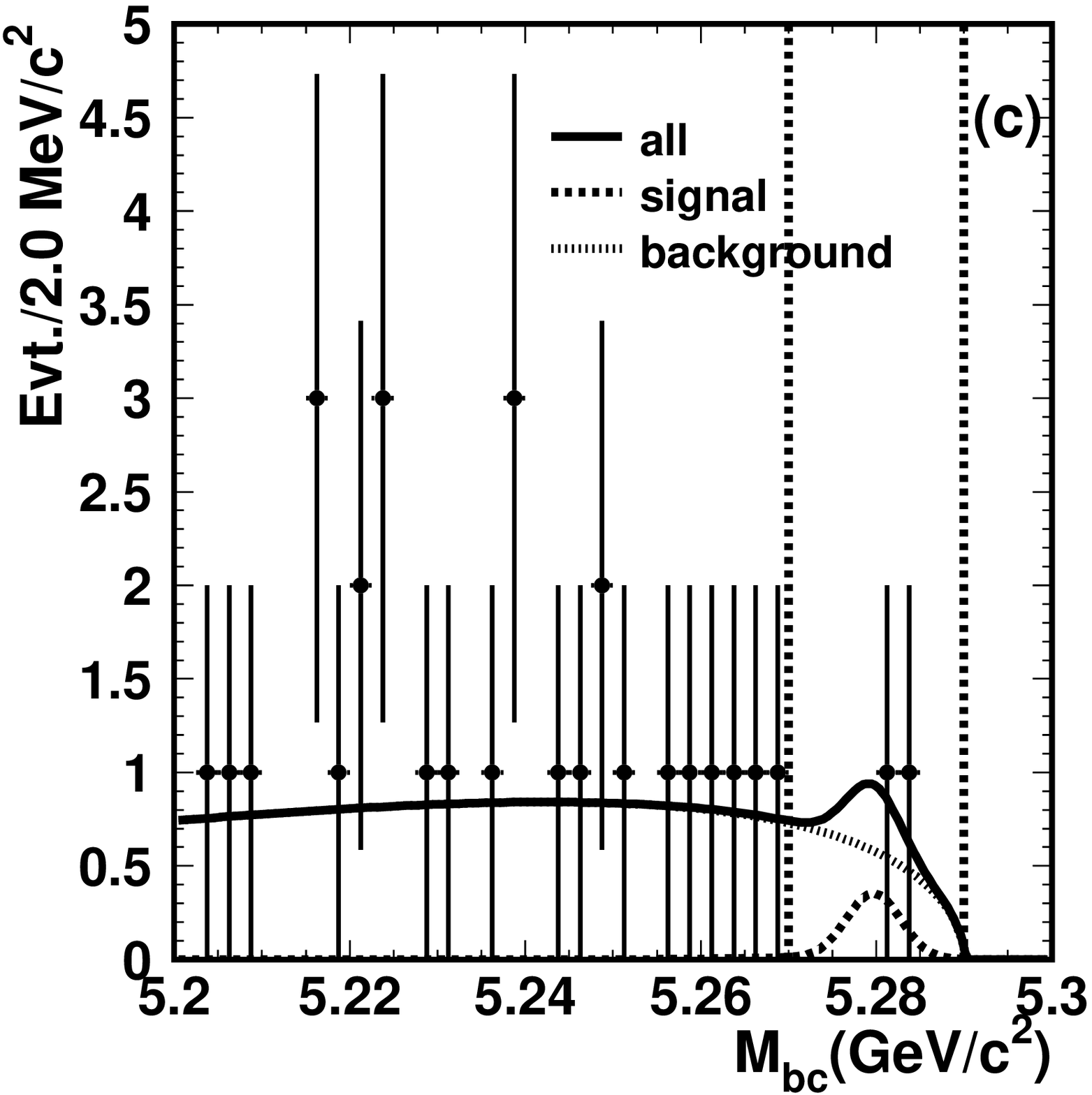}
\caption{ (a) The ($\Mbc$, $\DE$) scatterplot of $B^{0} \to J/\psi
\eta' K_S^0$ candidates and its projections onto (b) $\DE$ with 5.27
GeV/$c^2$ $< \Mbc <$ 5.29 GeV/$c^2$ and (c) $\Mbc$ with $ |\DE| <
0.03$ GeV. The dashed lines and the solid box indicate the signal
regions. The curves are the result of the fit. } \label{sbox2}
\end{figure*}

\begin{table*}[!hbtp]
\begin{center}
\caption{\label{result} Summary of the results. $Y$ and $b$ are the
signal and expected total background yields in the signal region,
Sig. is the statistical significance, $n_{0}$ is the observed number
of candidate events in the signal region, $Y_{90}$ is the upper
limit on the signal yield at 90\% confidence level, $\epsilon$
(error includes systematic error) is the detection efficiency and
${\cal B}$ is the branching fraction.}
\begin{tabular}{cccccccc}
\hline \hline
Mode & $Y$ &$b$ & Sig. & $n_0$ &$Y_{90}$ & $\epsilon(\%)$ & ${\cal B}$ \\
\hline $B^{+} \to J/\psi \eta ' K^+$ & ~$6.0^{+ 4.5}_{-3.9}$ &
~$22.0 \pm 0.7$ & ~$1.8\sigma$ &
~37 & ~$28.1$ & ~$3.9 \pm 0.6$ & ~~$< 8.8 \times 10^{-5}$ \\

$B^{0} \to J/\psi \eta ' K_S^0$  &$ 1.1^{+1.9}_{-1.2}$
 & $4.2 \pm 0.3$ & $0.9\sigma$ & 2
 & $3.7$ & $2.7_{-0.7}^{+0.6}$ & ~~$<2.5 \times 10^{-5}$ \\
\hline\hline
\end{tabular}
\end{center}
\end{table*}

\begin {table*}[htp]
\begin {center}
\caption {Summary of systematic uncertainties (\%) of the
denominator in branching fraction calculation.}
\begin {tabular}{ccc}
\hline \hline
Source&$B^+ \to J/\psi \eta' K^+$&$B^0 \to J/\psi \eta' K_S^0$\\
\hline
Tracking error & $\pm 7.4$ &$\pm 8.9$   \\
PID~($\pi$ and K) & $\pm 2.7$ & $\pm 1.4$ \\
Lepton ID & $\pm 3.8$ & $\pm 3.8$  \\
$\pi^0$ veto & $\pm 3.0$ & $\pm 3.0$ \\
$\eta$ reconstruction & $\pm 4.3$ & $\pm 4.3$ \\
$K_S^0$ reconstruction & - & $\pm 3.7$ \\
Branching fractions & $\pm 3.3$ & $\pm 3.3$ \\
MC statistics & $\pm 0.2 $ & $\pm 0.3 $  \\
3-body decay model & $+9.4$/$-11.3$ & $+17.5$/$-22.8$    \\
$N_{B\bar{B}}$ & $\pm 1.4$ & $\pm 1.4$ \\
\hline
Total & $+14.3$/$-15.6$ & $+21.4$/$-25.9$ \\
\hline \hline
\end {tabular}
\label{sys-err}
\end {center}
\end {table*}

In the fit, the value of $N_k$ and the parameters for the
combinatoric background are allowed to float. Figures \ref{sbox} and
\ref{sbox2} show the ($\Mbc$, $\DE$) scatterplots and their
projections for candidates after all selections are applied. The fit
results are superimposed on the projections. There are 37 candidate
events in the signal region for $B^{+} \to J/\psi \eta ' K^+$ and
two for $B^{0} \to J/\psi \eta ' K_S^0$.

Table~\ref{result} summarizes the maximum-likelihood fit results for
the signal ($Y$) and signal-region background ($b$) yields and their
statistical errors. The statistical significance is defined as
$\sqrt{-2\ln({\cal L}_{0}/{\cal L}_{\rm max})}$, where ${\cal
L}_{\rm max}$ and ${\cal L}_{0}$ denote the maximum likelihood with
the fitted signal yield and with the yield fixed at zero,
respectively.

No significant signal is found for either the $B^{+}\to J/\psi \eta
' K^+$ or $B^{0}\to J/\psi \eta ' K_S^0$ decay mode. We obtain upper
limits on the yield at 90\% confidence level ($Y_{90}$) from number
of observed candidates in the signal region ($n_0$) and $b$ using
the Feldman-Cousins method \cite{FC}. We account for systematic
uncertainty due to uncertainties in the signal reconstruction
efficiency and background estimate using the method of Ref.
\cite{Conrad}.

The branching fraction is determined with $N_{S}/[\epsilon \times
N_{B\bar{B}} \times {\cal B}(J/\psi \to l^{+}l^{-}) \times {\cal
B}(\eta' \to \eta \pi^+ \pi^-) \times {\cal B}(\eta \to \gamma
\gamma)]$ for $B^{+}\to J/\psi \eta ' K^+$ and $N_{S}/[\epsilon
\times N_{B\bar{B}}\times {\cal B}(J/\psi \to l^{+}l^{-}) \times
{\cal B}(\eta' \to \eta \pi^+ \pi^-) \times {\cal B}(\eta \to \gamma
\gamma) \times {\cal B}(K_S^0 \to \pi^+ \pi^-)]$ for $B^{0}\to
J/\psi \eta ' K_S^0$. Here $N_{S}$ is the signal yield,
$N_{B\bar{B}}$ is the number of $B\bar{B}$ pairs. We use the world
averages \cite{pdg} for the branching fractions of ${\cal B}(J/\psi
\to l^{+}l^{-})$, ${\cal B}(\eta' \to \eta \pi^+ \pi^-)$, ${\cal
B}(\eta \to \gamma \gamma)$ and ${\cal B}(K_S^0 \to \pi^+ \pi^-)$.
The efficiencies ($\epsilon$) are determined from the signal MC
sample with the same selection as used in the data. A three-body
phase space model is employed for all three decay modes. The
fractions of neutral and charged $B$ mesons produced in
$\Upsilon(4S)$ decays are assumed to be equal.

The sources and sizes of systematic uncertainties in branching
fraction calculation are summarized in Table~\ref{sys-err}. The
dominant sources are uncertainties in the three-body decay model,
tracking efficiency and particle identification. For the error due
to uncertainty in decay modeling for $B^+ \to J/\psi \eta ' K^+$
($B^0 \to J/\psi \eta ' K_S^0$), the distribution in phase space is
unknown. We conservatively assign the maximum variation of
efficiency among the slices of M($J/\psi$,$\eta '$),
M($J/\psi$,$K^-$) and M($\eta '$,$K^-$) [M($J/\psi$,$\eta '$),
M($J/\psi$,$K_S^0$) and M($\eta '$,$K_S^0$)] as a systematic
uncertainty. The uncertainties in the tracking efficiency are
estimated by linearly adding the momentum-dependent single track
systematic errors. We assign uncertainties of about 1.3\% per kaon,
about 0.7\% per pion, and about 1.9\% per lepton for the particle
identification. The $\eta$ reconstruction uncertainty is determined
by measuring the efficiency ratio between data and MC sample in the
inclusive $\eta$ sample. The $K^0_S$ reconstruction is checked by
comparing the ratio of $D^+ \to K^0_S \pi^+$ and $D^+ \to K^-
\pi^+\pi^-$ yields.

The systematic errors for the background yield are evaluated by
varying each of the PDF parameters by its statistical error from the
fit.

In summary, we searched for $B^{+} \to J/\psi \eta' K^+$ and $B^{0}
\to J/\psi \eta' K_S^0$ decays. No statistically significant signal
was found for either of the two decay modes; upper limits for the
branching fractions are determined to be $\mathcal{B}(B^{+} \to
J/\psi \eta' K^+) < 8.8 \times 10^{-5}$ and $\mathcal{B}(B^{0} \to
J/\psi \eta' K_S^0) < 2.5 \times 10^{-5}$ at 90\% confidence level.

\begin{acknowledgments}
We thank the KEKB group for the excellent operation of the
accelerator, the KEK cryogenics group for the efficient operation of
the solenoid, and the KEK computer group and the NII for valuable
computing and Super-SINET network support.  We acknowledge support
from MEXT and JSPS (Japan); ARC and DEST (Australia); NSFC and KIP
of CAS (contract No. 10575109 and IHEP-U-503, China); DST (India);
the BK21 program of MOEHRD, and the CHEP SRC and BR (grant No.
R01-2005-000-10089-0) programs of KOSEF (Korea); KBN (contract
No.~2P03B 01324, Poland); MIST (Russia); MHEST (Slovenia);  SNSF
(Switzerland); NSC and MOE (Taiwan); and DOE (USA).

\end{acknowledgments}


\begin{thebibliography}{999}
\bibitem{CC} Inclusion of the charge conjugate state is implied throughout
             this paper.
\end{thebibliography}
\end{document}